\documentclass[aps,prx,reprint,superscriptaddress]{revtex4-2}
\usepackage{graphicx}  
\usepackage{dcolumn}   
\usepackage{bm}        
\usepackage{bbold}
\usepackage{amssymb, amsmath}   
\usepackage[usenames, dvipsnames]{color}
\usepackage{float}
\usepackage{pythonhighlight}
\usepackage{colortbl}
\usepackage{braket}
\usepackage{hyperref} 
\hypersetup{
	unicode=true,                                           
	plainpages=false,
	pdffitwindow=true,                                      
	colorlinks=true,                                        
	linkcolor=blue,                                        
	citecolor=blue,                                         
	filecolor=blue,                                        
	urlcolor=blue                                          
}
\usepackage{array, multirow}

\renewcommand\bra[1]{{\!\langle{#1}|}}
\renewcommand\ket[1]{{|{#1}\rangle}}

\begin{document}
\title{Non-perturbative exciton transfer rate analysis of the Fenna-Matthews-Olson photosynthetic complex under reduced and oxidised conditions}

\author{Hallmann Ó. Gestsson}%
\thanks{These authors contributed equally.}
\affiliation{Department of Physics and Astronomy, University College London, Gower Street, WC1E 6BT, London, United Kingdom}%
\author{Charlie Nation}%
\thanks{These authors contributed equally.}
\affiliation{Department of Physics and Astronomy, University College London, Gower Street, WC1E 6BT, London, United Kingdom}%
\author{Jacob S. Higgins}
\affiliation{JILA, National Institute of Standards and Technology and Department of Physics, University of Colorado, Boulder, CO, USA}%
\author{Gregory S. Engel}
\affiliation{Department of Chemistry, University of Chicago, Chicago, IL, 60637}%
\author{Alexandra Olaya-Castro}
\email{a.olaya@ucl.ac.uk}
\affiliation{Department of Physics and Astronomy, University College London, Gower Street, WC1E 6BT, London, United Kingdom}%
\date{\today}

\begin{abstract}
Two-dimensional optical spectroscopy experiments have shown that exciton transfer pathways in the Fenna-Matthews-Olson (FMO) photosynthetic complex differ drastically under reduced and oxidised conditions, suggesting a functional role for collective vibronic mechanisms that may be active in the reduced form but attenuated in the oxidised state. Higgins \textit{et al.} [PNAS 118 (11) e2018240118 (2021)] used Redfield theory to link the experimental observations to altered exciton transfer rates due to oxidative onsite energy shifts that detune excitonic energy gaps from a specific vibrational frequency of the bacteriochlorophyll \textit{a}.
Using a memory kernel formulation of the hierarchical equations of motion, we present non-perturbative estimations of transfer rates that yield a modified physical picture. Our findings indicate that onsite energy shifts alone cannot reproduce the observed rate changes in oxidative environments, either qualitatively or quantitatively. By systematically examining combined changes both in site energies and the local environment for the oxidised complex, while maintaining consistency with absorption spectra, our results suggest that vibronic tuning of transfer rates may indeed be active in the reduced complex. However, we achieve qualitative, but not quantitative, agreement with the experimentally measured rates.  Our analysis indicates potential limitations of the FMO electronic Hamiltonian, which was originally derived by fitting spectra to second-order cumulant and Redfield theories. This suggests that reassessment of these electronic parameters with a non-perturbative scheme, or derived from first principles, is essential for a consistent and accurate understanding of exciton dynamics in FMO under varying redox conditions.
\end{abstract}

\maketitle

\section{Introduction}
The Fenna-Matthews-Olson (FMO) protein \cite{Olson1962Jun, Fenna1975Dec} found in green sulfur bacteria \textit{Chlorobaculum tepidum} \cite{Wahlund1991Aug} is a homotrimer whose subunits each consist of eight interacting bacteriochlorophyll (BChl) \textit{a} that collectively absorb light and facilitate excitation energy transfer (EET) in the form of delocalised excitons to the reaction centre \cite{vanAmerongen2000Jun, Hauska2001Oct, Olson2004Apr, SchmidtamBusch2011Jan, May2011Feb}. 
It has become an exemplary system for the study of EET in photosynthetic complexes \cite{Adolphs2006Oct, Ishizaki2009Oct, Ritschel2011Nov, Dostal2016Jul, Kell2016Aug, Sokolov2024Jul, Dare2024Oct}, particularly since reports of long-lived quantum beats in two-dimensional electronic spectroscopy (2DES) both at low and physiological temperatures \cite{Engel2007Apr, Hayes2010Jun, Panitchayangkoon2010Jul, Fidler2012Jul}. 
These findings suggest that non-trivial quantum phenomena may play an important role in the biological function of the FMO complex, but there is no yet conclusive evidence or consensus regarding their significance \cite{Duan2017Aug}. 

A central hypothesis for the observed beats in the FMO, and in similar studies for other complexes \cite{Richards2012Jan, Christensson2012Jun, Kolli2012Nov, Chin2013Feb, O'Reilly2014Jan, Dean2016Dec, Blau2018Apr, Caycedo-Soler2018Jun, Arsenault2020Mar}, is the presence of vibronic mechanisms whereby a specific vibrational motion couples to excited state dynamics acting to sustain coherence and assist energy transfer. 
The recent discovery of redox-sensitive cysteines in FMO by Orf \textit{et al.} \cite{Orf2016Aug} has enabled a new approach to investigate this hypothesis and to further understand the possible importance of such a mechanism in biological function. 
To this end, 2DES experiments have investigated the energy transfer dynamics in FMO proteins under both reducing and oxidising conditions \cite{Allodi2018Jan, Higgins2021Mar, Higgins2021Dec}. 
These studies have revealed some key significant features: quantum beats observed in the reduced FMO complex disappear under oxidised conditions \cite{Higgins2021Dec}, and specific exciton transfer pathways are significantly attenuated in the latter \cite{Allodi2018Jan, Higgins2021Mar}. 
Based on this, Higgins \textit{et al.} \cite{Higgins2021Mar} posit the modulation of vibronic processes as key to the redox mechanism of FMO, acting to funnel excitations towards quenching sites in oxidising conditions and stemming the generation of reactive oxygen species that can damage the reaction centre. 

However, in the above experiments the hypothesis was rationalised within the Redfield theoretical framework \cite{Redfield1957Jan, Redfield1965Jan, Pollard1994Apr, Ishizaki2009JunPartI} which is known to have limited applicability \cite{Ishizaki2009JunPartI}. 
An important question is whether the energy transfer rate analysis and hypothesis that have been put forward for the observed differences in reduced and oxidised FMO in the above works hold if we apply a more accurate framework, namely the hierarchical equations of motion (HEOM) \cite{Tanimura1989Jan, Ishizaki2005Dec, Ishizaki2009JunPartII, Shi2009Feb, Tanaka2009Jul, Tanimura2020Jul}. 
This is not a simple question to address as the concept of transfer rate is not clear when the dynamics is non-Markovian and multi-exponential. 
Recent theory has addressed this issue by recasting the set of HEOM to a memory kernel form to define generalised transfer rates \cite{Jesenko2013May, Jesenko2014May, Zhang2016May}. 
In this work, we apply this framework to develop a systematic comparison of the experimental and predicted exciton kinetics according to both HEOM and Redfield, to assess whether the proposed modulation of vibronic processes holds when an appropriate non-perturbative treatment of the vibrational environment is considered. 
We observe a stark disagreement between the predicted HEOM and Redfield excitonic transfer rates for spectral densities that include an underdamped vibrational motion resonant to an excitonic transition in the FMO. 
This leads us to conclude that perturbative approaches to describe effective transfer rates in systems such as FMO are liable to lead to inconsistencies with experimentally determined rates via ultrafast spectroscopy; therefore efforts to rationalise experimental observations of exciton kinetics in such systems should rely on non-perturbative approaches such as HEOM \cite{Tanimura1989Jan, Ishizaki2005Dec, Ishizaki2009JunPartII, Shi2009Feb, Tanaka2009Jul, Tanimura2020Jul}, tensor-network models \cite{Pollock2018Jan, Strathearn2018Aug, Fux2023Aug, Cygorek2024Feb}, and direct path-integral approaches \cite{Makri1995Mar, Makri2020Jul}. 

Furthermore, we demonstrate that the model proposed by Higgins \textit{et al.} \cite{Higgins2021Mar} for the oxidised FMO does not reproduce experimental measurements of the absorption spectra \cite{Allodi2018Jan}, prompting us to develop an investigation of generalised excitonic transfer rates over a parameter range consistent with linear spectra observations. 
The absorption spectrum is a key fingerprint of a photosynthetic complex, and any theoretical model must first and foremost be able to recover it. 
By restricting the variation of the electronic and vibrational parameters of the oxidised model to be faithful to its corresponding absorption spectrum we are unable to simultaneously recover the observed EET rates. 
We suggest that a potential explanation for this discrepancy is that the employed base electronic Hamiltonian model, which was generated by Kell \textit{et al.} \cite{Kell2016Aug} via fitting experimental observations with a second-order cumulant approach and Redfield theory, may be unsuitable for reducing conditions; this is indicated by the fact that, when considering the accurate framework it does not lead to generalised transfer rates that compare with experimental rates for important transfer pathways, even upon the inclusion of a relevant underdamped vibrational motion. 
Therefore, our work suggests that further theoretical analyses that include a re-evaluation and more accurate determination of the departing electronic Hamiltonian parameters are needed to understand the changes observed in the exciton dynamics in FMO under reduced and oxidised conditions.

The structure of this paper is as follows: in Section \ref{sec:model} we introduce the Frenkel exciton model and the vibrational bath modes that simulate the FMO complex, and then outline an approach to computing EET rates from HEOM in Section \ref{sec:rates_theory}. 
In Section \ref{sec:tuning_paper_analysis} we demonstrate that the previously proposed model of oxidised FMO does not reproduce the experimental absorption spectra and furthermore show that the Redfield approach to rate derivation is inconsistent with HEOM predictions.
For Section \ref{sec:env_model} we show that a non-perturbative description of the environment yields significantly different qualitative behaviour of the exciton rates compared to Redfield theory when seeking an alternative model. 
In Section \ref{sec:redox_mechanism} we analyse the behaviour of pertinent transfer rates when varying relevant site energies. 
The conclusions and outlook are then presented in Section \ref{sec:conclusions}. 

\begin{figure}
    \centering
    \includegraphics[width=\linewidth]{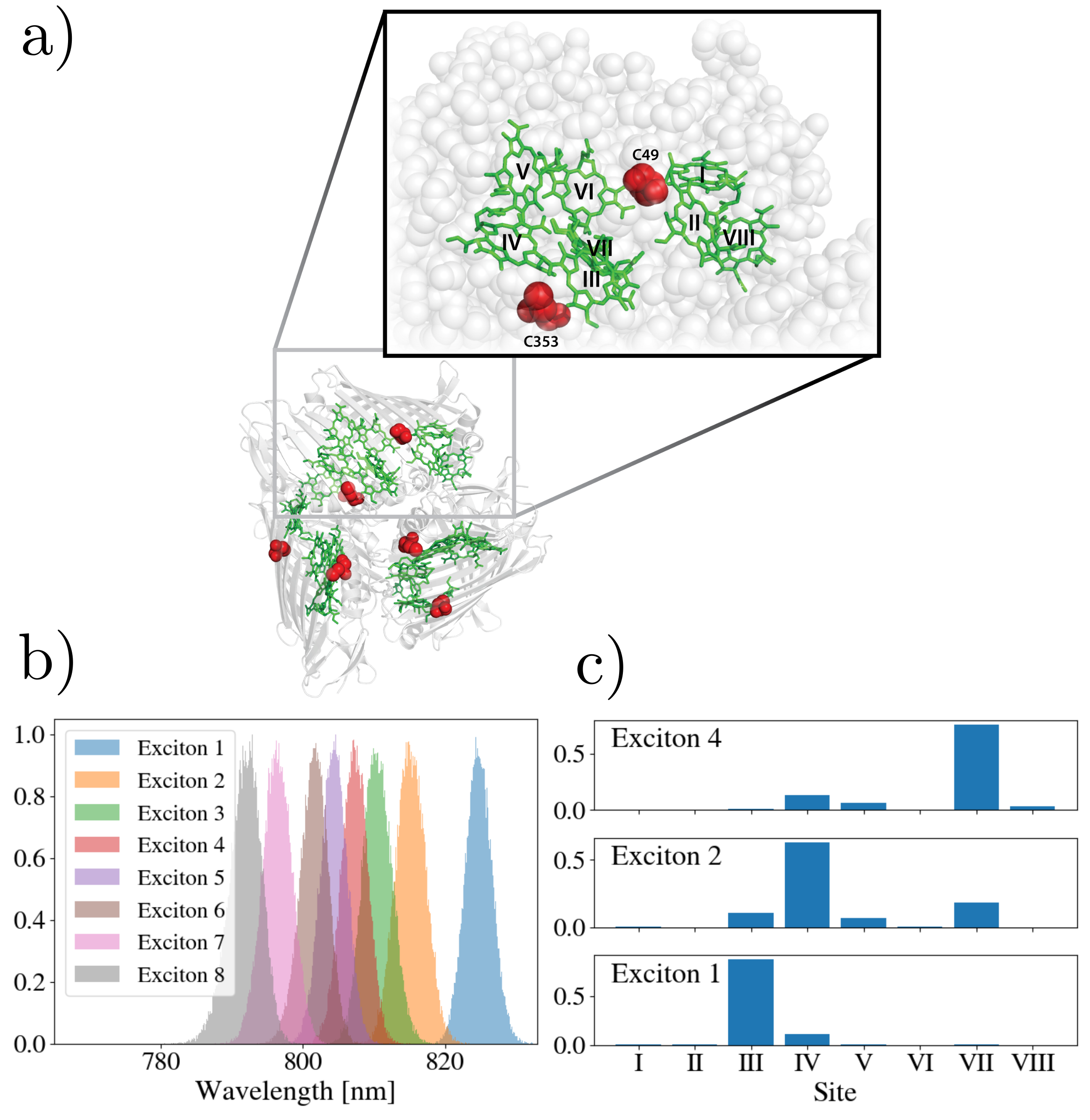}
    \caption{a) The Fenna-Matthews-Olson (FMO) complex, sites labelled I-VIII, C49 and C353 label the side chains of cysteines responsible for excitation quenching \cite{Orf2016Aug} b) Shows the distribution of excitonic energies (in terms of their corresponding wavelength) for 100000 realisations of the disordered Hamiltonian. c) Shows the site participations for excitons 4, 2, and 1 with no average over disorder.}
    \label{fig:FMO}
\end{figure}

\section{Model}\label{sec:model}
To simulate the EET dynamics for FMO, we use a Frenkel exciton Hamiltonian of the form (we set $\hbar=1$ throughout the study)
\begin{equation}\label{eq:H_S}
    \hat{H}_S = \sum_{i=1}^8 \epsilon_i \ket{i}\bra{i} + \sum_{i\neq j} V_{ij}\ket{i}\bra{j},
\end{equation}
where $\ket{i}$ is the excited state of the $i$-th BChl while others are in the ground state, and $\epsilon_i$ and $V_{ij}$ are BChl site energies and interaction strengths, respectively \cite{vanAmerongen2000Jun, May2011Feb}. 
We denote the energy eigenstates or exciton states of $\hat{H}_S$ as $|\alpha\rangle$ with eigenenergy $E_{\alpha}$. 
Hereafter, we number the BChl sites using Roman numerals I $-$ VIII and the exciton states using Arabic numerals $1-8$. 
For meaningful comparison with the report in Ref.\ \cite{Higgins2021Mar}, we considered the electronic Hamiltonian generated by Kell \textit{et al.} \cite{Kell2016Aug} that is referred to as `model C.' 
Appendix \ref{app:FMO} summarises the parameters of this model.  
As in `model C,' we consider static disorder for the on-site energies given by independent Gaussian distributions with  full width at half maximum of 125 cm$^{-1}$ for all sites, except for site III for which it is 75 cm$^{-1}$. 
In Fig.\ \ref{fig:FMO} we present a diagram of the structure of FMO alongside excitonic energy distributions and site participations, i.e. $|\langle i|\alpha\rangle|^2$ in relevant excitons.

The bath is modelled by a set of harmonic oscillators, $\hat{H}_B = \sum_{i, k} \omega_{k, i} \hat{b}^\dagger_{k, i} \hat{b}_{k, i}$, whose displacements are linearly coupled with coupling strength $g_{k,i}$ to its corresponding local BChl population as $\hat{H}_{SB} = \sum_i \ket{i}\bra{i} \sum_k g_{k, i} (\hat{b}_{k, i} + \hat{b}^\dagger_{k, i})$ where $\hat{b}^\dagger_{k, i}$ $(\hat{b}_{k, i})$ are creation (annihilation) of local bath mode $k$ of site $i$. 
Interaction strength of BChls to their respective local vibrational environments is therefore quantified via a spectral density composed of a low-energy phonon background and an underdamped mode as
\begin{equation}
    J_i(\omega) = J_{i, p}(\omega) + J_u(\omega),
\end{equation}
where we have a site-dependent background $J_{i, p}(\omega)$ and a site-independent mode $J_u(\omega)$. The models \cite{Kell2016Aug, Higgins2021Mar} use a log-normal distribution for the spectral density, which is not a tractable form for the HEOM. Therefore, we make use of a best-fit to an underdamped Brownian oscillator (UBO) (with reorganisation energy $\lambda$, damping rate $\gamma$, and peak position $\Omega$) whose form,
\begin{equation}\label{eq:ubo_spectral_density}
    J(\omega) = 2\lambda\gamma\Omega^2\frac{\omega}{(\omega^2-\Omega^2)^2 + \gamma^2\omega^2},
\end{equation}
is amenable to HEOM, and also allows us to fully reproduce the theoretical rates calculated using Redfield theory \cite{Adolphs2006Oct, Kell2016Aug, Higgins2021Mar}. Therefore, we are in a position to make a fair comparison of the exciton transfer rates computed using Redfield and HEOM. 

\begin{figure}
    \centering
    \includegraphics[width=\linewidth]{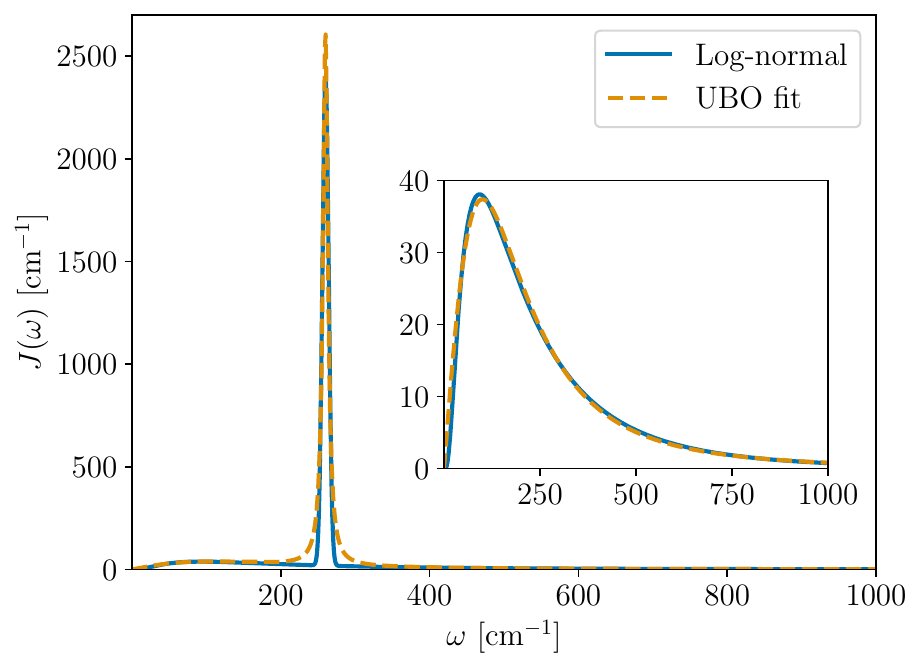}
    \caption{Comparison of the spectral density composed of a log-normal and underdamped mode to a best-fit UBO model. Inset demonstrates the fit of the single UBO to the Log-normal component. The quantitative agreement between the two allows us to make a fair comparison of rates computed using Redfield and HEOM. }
    \label{fig:ln_vs_fit}
\end{figure}

Fitting the functional form of Eq.\ (\ref{eq:ubo_spectral_density}) for all sites, except for site III, the background component with parameters $\lambda_{i,p} = 30$ cm$^{-1}$, $\gamma_{i,p} = 415$ cm$^{-1}$, $\Omega_{i,p} = 190$ cm$^{-1}$ for $i\neq3$. The background for site III is distinct such that we have $\lambda_{3,p} = 23$ cm$^{-1}$, $\gamma_{3,p} = 165$ cm$^{-1}$, $\Omega_{3,p} = 100$ cm$^{-1}$. For the underdamped mode, $\lambda_u = 40$ cm$^{-1}$, $\gamma_u = 8$ cm$^{-1}$, $\Omega_u = 260$ cm$^{-1}$. We demonstrate that the fit is faithful to the original spectral density with a side-by-side comparison, as shown in Fig.\ \ref{fig:ln_vs_fit}. All the numerical results shown throughout the paper assume a vibrational environment at a temperature of 77K. 

We use the following expression for absorption spectra,
\begin{equation}
    A(\omega) = 2\sum_{p=x,y,z}\text{Re}\left[\int_0^\infty\text{d}t \text{Tr}(\hat{\mu}_p e^{\mathcal{L}t}\left[\hat{\mu}_p\rho_0\right])e^{i\omega t}\right],
\end{equation}
where $\mathcal{L}$ is the dynamical generator of the irreversible reduced system dynamics for Frenkel excitons, and $\hat{\mu}_p = \sum_i \mu_{p,i}\ket{i}\bra{0} + \text{h.c.}$ is the total transition dipole for a given polarisation and the average is taken with respect to the ground state, $\rho_0=\ket{0}\bra{0}$. Structural data pertaining to the transition dipoles (the $\mu_{p,i}$ elements) were taken from the Protein Data Bank (code: 3ENI) \cite{Tronrud2009May}. See Appendix \ref{app:linear_absorption} for further computational details on the linear spectra. 

\section{Non-perturbative exciton transfer rates}\label{sec:rates_theory}
We now address the issue of defining the effective exciton transfer rates, which are necessary for understanding the quenching mechanisms. 
In the context of Redfield theory, the transfer rates between exciton states are well-defined and are given by the diagonal elements of the Redfield tensor (see Appendix \ref{app:redfield}). 
However, it is well known that Redfield and other perturbative theories may not accurately describe excited state dynamics for photosynthetic complexes, and non-perturbative methods are called for \cite{Ishizaki2009JunPartI}.
HEOM is a framework capable of accurately describing exciton dynamics in vibrational environments that approximates those of photosynthetic systems;
unlike Redfield, there are no clearly defined rates in HEOM associated with the exciton-to-exciton population transfer. 
This is effectively due to the fact that in order to account for the environmental non-Markovian influence the resulting equation of motion for the reduced system is not local in time. 
In this section, we utilise projection operator techniques to extract the transfer rates from the memory kernel reformalisation of HEOM \cite{Nakajima1958Dec, Zwanzig1960Nov, Jesenko2013May, Jesenko2014May, Zhang2016May}. 

For HEOM we assume a factorised initial state of the form $\rho(0) = \rho_S(0)\otimes\rho_B(0)$, where $\rho_B(0) = e^{-\beta\hat{H}_B} / \text{Tr}(e^{-\beta\hat{H}_B})$ is a thermalised state of the bath at an inverse temperature $\beta$, and further expand the environmental correlation functions in terms of exponential functions as
\begin{equation}
    C_i(t) = \sum_{k}(c_{i,k}^+e^{-\nu_{i,k}^+ t} + c_{i,k}^-e^{-\nu_{i,k}^- t}) + \sum_{k^\prime} c_{i,k^\prime}e^{-\nu_{i,k^\prime} t},
\end{equation}
where we have complex conjugate pairs of rates ($(\nu_{i,k}^+)^* = \nu_{i,k}^-$), and real-valued rates. The expressions for the coefficients and rates of the expansion are provided in Appendix \ref{app:heom}. 
The reduced system dynamics may be expanded into an infinite hierarchy of dynamically coupled auxiliary density operators as
\begin{equation}\label{eq:HEOM_infinite}
   \partial_t \rho_{\vec{n}} = (-i\hat{H}_S^\times - \sum_{r\in \mathcal{S}} n_{r} \nu_{r} )\rho_{\vec{n}} + \sum_{r\in \mathcal{S}} \left(\hat{n}_i^\times \rho_{\vec{n}^+_{r}} - n_{r} \hat{\Theta}_{r} \rho_{\vec{n}^-_{r}}\right),
\end{equation}
where the elements of $\mathcal{S}$ are of the form $\{i,k,\sigma\}$ with $i$ and $k$ indexing BChl sites and the terms of $C_i(t)$, respectively. 
We also have $\sigma=\pm$ to index the conjugate pairs of rates, and drop it for real-valued rates. 
The auxiliary indices $\vec{n}^\pm_{r}$ denote the addition of $\pm1$ at the $r$-th entry of $\vec{n}$ such that super-operators $\hat{n}_i^\times$ and $\hat{\Theta}_{r}$ are responsible for couplings between tiers in the hierarchy. 
The $\times$ superscript denotes a commutator and we have set $\hat{n}_i = \ket{i}\bra{i}$, 
\begin{equation}
    \hat{\Theta}_{i,k,\pm} = \frac{c_{i,k}^\pm + (c_{i,k}^\mp)^*}{2}\hat{n}_i^\times + \frac{c_{i,k}^\pm - (c_{i,k}^\mp)^*}{2}\hat{n}_i^\circ,
\end{equation}
\begin{equation}
    \hat{\Theta}_{i,k} = \text{Re}(c_{i,k})\hat{n}_i^\times + i\text{Im}(c_{i,k})\hat{n}_i^\circ,
\end{equation}
where the superscript $\circ$ denotes an anti-commutator.

We now represent the hierarchy as $\mathcal{L}$ such that the exact dynamics of the reduced density matrix $\rho_S$ of the system are generated by the equation
\begin{equation}
    \partial_t\rho = \mathcal{L}\rho,
\end{equation}
where $\rho = \left(\rho_S, \{\rho_{\Vec{n}};\Vec{n}\in A\}\right)^T$ is a column vector consisting of $\rho_S$ and all auxiliary density operators, with $A$ as the set of auxiliary indices. 
We then define a projection operator $\mathcal{P}$ whose action on $\rho$ is given as $\mathcal{P}\rho = \left(\rho_S, \{0;\Vec{n}\in A\}\right)^T$, and its complement $\mathcal{Q} = \mathcal{I} - \mathcal{P}$ such that $\mathcal{Q}\rho = \left(0, \{\rho_{\Vec{n}};\Vec{n}\in A\}\right)^T$, where $\mathcal{I}$ is the identity operator. This results in the following coupled set of equations
\begin{align}\label{eq:NaZw_projection_equation}
    \partial_t\mathcal{P}\rho &= \mathcal{PLP}\rho + \mathcal{PLQ}\rho, \\
    \partial_t\mathcal{Q}\rho &= \mathcal{QLP}\rho + \mathcal{QLQ}\rho,
\end{align}
for which the formal solution of $\mathcal{Q}\rho$ is
\begin{equation}
    \mathcal{Q}\rho(t) = e^{\mathcal{QL}t}\left(\mathcal{Q}\rho(0) - \int_0^t\text{d}\tau e^{-\mathcal{QL}\tau}\mathcal{QLP}\rho(\tau)\right),
\end{equation}
which then is inserted into Eq.\ \eqref{eq:NaZw_projection_equation} in order to obtain
\begin{equation}\label{eq:genQME}
    \partial_t\mathcal{P}\rho(t)= \mathcal{PLP}\rho(t) + \int_0^t\text{d}\tau \mathcal{PL}e^{\mathcal{QL}(t-\tau)}\mathcal{QLP}\rho(\tau).
\end{equation}
The first term corresponds to the unitary dynamics of the reduced system ($\mathcal{PLP}\rho = \left(-i\left[\hat{H}_S, \rho_S\right], \{0;\Vec{n}\in A\}\right)^T$). 
The inhomogeneous term does not contribute to the dynamics, as we assumed a factorised initial state for the system and environment, with the environment in a thermal state. 
The final term captures the influence of the environment on the reduced system dynamics in terms of a memory kernel, which can be identified as
\begin{equation}
    \mathcal{K}(t) = \mathcal{PL}e^{\mathcal{QL}t}\mathcal{QL},
\end{equation}
such that for our purposes we now have an integro-differential equation of motion for the reduced density matrix dynamics
\begin{equation}\label{eq:memory_kernel_equation}
    \partial_t\mathcal{P}\rho(t) = \mathcal{PLP}\rho(t) + \int_0^t\text{d}\tau \mathcal{K}(t-\tau)\mathcal{P}\rho(\tau).
\end{equation}

Equation (\ref{eq:memory_kernel_equation}) is rendered to a quantum master equation in the long-time limit such that we can extract the exact EET rates of the non-equilibrium steady state that are directly analogous to rates stemming from Redfield theory. 
In the long-time limit the environmental component becomes
\begin{equation}
    \int_0^\infty\text{d}\tau \mathcal{K}(\tau) = \mathcal{PLGQL},
\end{equation}
where we set $\mathcal{G} = -\lim_{\epsilon\rightarrow 0^+} \frac{1}{\mathcal{QL} - \epsilon\mathcal{I}}$. We thereby recover the operator from which we can extract exciton transfer rates by expanding in the eigenstate basis.
Given eigenstates $\ket{\alpha}$, $\ket{\beta}$ of $\hat{H}_S$ we may numerically compute the exact population transfer rate from $\ket{\alpha}$ to $\ket{\beta}$ as
\begin{equation}\label{eq:long_time_limit_rate}
    W_{\alpha\to\beta} = \braket{\beta\vert\mathcal{PLGQLP}\left[\ket{\alpha}\bra{\alpha}\right]\vert\beta}.
\end{equation}
We include the Ishizaki-Tanimura correction term \cite{Ishizaki2005Dec} in the definition of the rate in Eq.\ \eqref{eq:long_time_limit_rate}, and compute
\begin{equation}
   W_{\alpha\to\beta} = \braket{\beta\vert(\mathcal{P}\Xi\mathcal{P} + \mathcal{PLGQLP})\left[\ket{\alpha}\bra{\alpha}\right]\vert\beta},
\end{equation}
with $\Xi = \sum_i \hat{n}_i^\times(\frac{1}{\beta}\partial_\omega J_i(\omega)\vert_{\omega=0}\hat{n}_i^\times - i\lambda_i\hat{n}_i^\circ - \sum_\sigma\sum_{k=0}^M \frac{1}{\nu_{i,k}^\sigma}\hat{\Theta}_{i,k,\sigma})$ where $M$ is the number of Matsubara terms accounted for by the hierarchy.
The reason for this amendment to the rates comes from the fact that we wish to account for every incoherent process, which would include Lindblad terms that this correction term represents. 
Finally, we note that we truncate the hierarchy by setting $\rho_{\vec{n}}$ with indices $\vec{n}$ that satisfy $\sum_{r\in\mathcal{S}} n_r > T$, for a pre-set threshold $T$, to zero, and that we utilise the scaling proposed by Shi \textit{et al.} \cite{Shi2009Feb} to improve the numerical efficiency of the HEOM method.

\begin{figure*}
    \includegraphics[width=\textwidth]{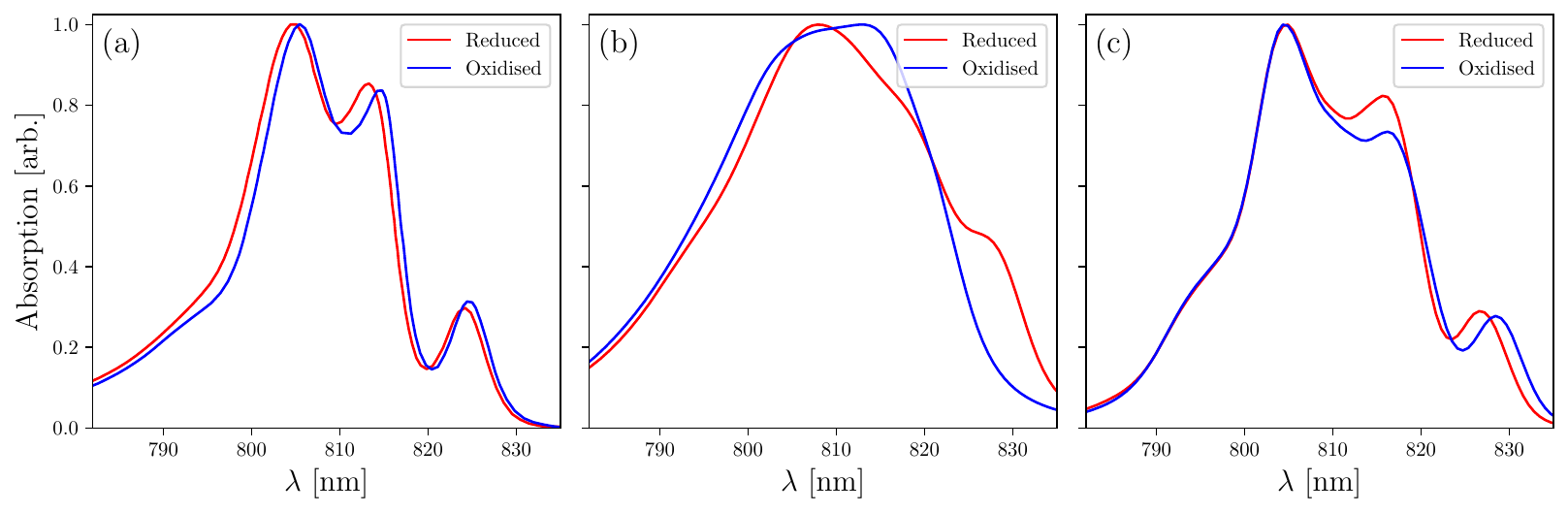}
    \caption{Comparison of the a) experimentally observed linear spectra with b) calculated linear spectra with HEOM for FMO model C \cite{Kell2016Aug}, site shift model of oxidation with shifts taken from \cite{Higgins2021Mar} of $\Delta\epsilon_2 =$ 40 cm$^{-1}$, $\Delta\epsilon_3 = $ 120 cm$^{-1}$, and $\Delta\epsilon_4 =$ 70 cm$^{-1}$ for oxidised model, with fitted two-UBO environment described in Section \ref{sec:model}, and c) AE model, which consists of model C Hamiltonian with site energy shifts of $\Delta\epsilon_2 = - 10$ cm$^{-1}$, $\Delta\epsilon_3 = - 25$ cm$^{-1}$, and $\Delta\epsilon_4 =- 25$ cm$^{-1}$, with Drude-Lorentz background environment (see Section \ref{sec:env_model}) and UBO mode centred at $260$ cm$^{-1}$ at $150$ cm$^{-1}$, for reduced and oxidised conditions respectively.}
    \label{Fig:linear_spectra}
\end{figure*}

\begin{table*}
    \caption{Exciton transfer rates in oxidised and reduced conditions. Experimental rates taken from Higgins et al. \cite{Higgins2021Mar} as inverse of measured timescales. This is compared to the model (labelled PNAS) of Higgins et al. \cite{Higgins2021Mar} calculated with Redfield and HEOM. We see significant differences in transfer rates between these approaches. Model labelled AE (altered environment) combines small site energy shifts (-10 cm$^{-1}$, -25 cm$^{-1}$, -25 cm$^{-1}$ when oxidised for sites II, III and IV, respectively) with a change in UBO mode frequency (from 260 cm$^{-1}$ in the reduced model to 150 cm$^{-1}$ in the oxidised form). In this case we capture the phenomenology of the experimental rates under oxidation but do not reproduce the magnitude. Each entry for Redfield and HEOM rates corresponds to the average of 30000 realisations of disorder.}
    \begin{tabular}{|c|c|c|c|c|c|c|c|c|}
        \hline
         & \multicolumn{2}{c|}{Experiment} & \multicolumn{2}{c|}{Redfield (PNAS)} & \multicolumn{2}{c|}{HEOM (PNAS)} & \multicolumn{2}{c|}{HEOM (AE)} \\ \hline
         Rate [ps$^{-1}$] & Reduced & Oxidised & Reduced & Oxidised & Reduced & Oxidised & Reduced & Oxidised \\ \hline
         $4\rightarrow 1$ & 1.98 & 0.67 & 2.51 & 0.92 & 0.67 & 2.01 & 0.45 & 0.30 \\
         $4\rightarrow 2$ & 2.45 & 4.41 & 2.17 & 2.11 & 4.86 & 4.09 & 2.03 & 2.73 \\
         $2\rightarrow 1$ & 2.20 & 2.27 & 2.24 & 2.80 & 4.41 & 4.37 & 2.60 & 1.98 \\ \hline
    \end{tabular}\label{tab:main_rates_table}
\end{table*}

\section{Inconsistency of proposed model with experimental absorption spectra and differences between HEOM and Redfield transfer rates}\label{sec:tuning_paper_analysis}
Herein, we summarise the findings of Higgins \textit{et al.} \cite{Higgins2021Mar} which demonstrates how the redox mechanism for FMO manifests as a change in exciton transfer dynamics. 
The central observation inferred from the experimental 2DES results is that the two redox-sensitive cysteines steer energy transfer in response to redox conditions. 
Under reducing conditions the transfer rate of the $4\to1$ pathway is similar to that of $4\to2$, and was determined to be $2.0$ ps$^{-1}$ and $2.5$ ps$^{-1}$, respectively. 
An excitation is therefore relatively unbiased towards either pathway. 
A drastic change in the two transfer rates was observed when oxidising conditions were induced. 
The $4\to1$ transfer rate decreases to $0.7$ ps$^{-1}$ and the $4\to2$ transfer rate increases to $4.4$ ps$^{-1}$, resulting in an effective propensity for the indirect $4\to2\to1$ EET pathway. 
It is proposed that this occurs due to an underdamped vibrational mode that for reducing conditions is resonant to the $4\rightarrow 1$ transition, which is then brought out of resonance by changes to the electronic parameters of the complex whenever oxidising conditions are encountered \cite{Higgins2021Mar}.
A Redfield model was used to rationalise the redox response, which manifests in the model as a shift to the energies of sites II, III, IV by $\Delta\epsilon_2 =$ 40 cm$^{-1}$, $\Delta\epsilon_3 = $ 120 cm$^{-1}$, $\Delta\epsilon_4 =$ 70 cm$^{-1}$, respectively, thereby altering the vibrational resonance conditions that in turn changes transfer rates within the system. 
The relevant experimentally determined transfer rates for the reduced and oxidised conditions are collated in Table \ref{tab:main_rates_table}, alongside their theoretical counterparts. 

\begin{figure*}[ht]
    \centering
    \includegraphics[width=\linewidth]{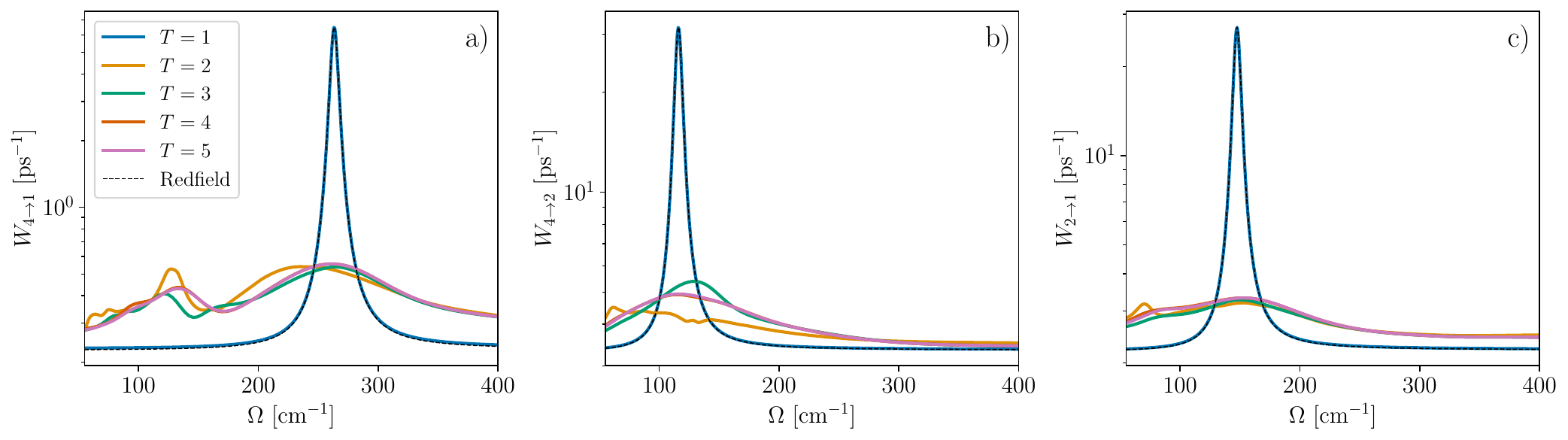}
    \caption{HEOM and Redfield rates $W_{4\to1}$, $W_{4\to2}$, $W_{2\to1}$ shown in log scale as a function of the UBO position $\Omega$. 
    The UBO mode has a fixed reorganisation energy $\lambda_u = 10$ cm$^{-1}$ and damping $\gamma_u = 8$ cm$^{-1}$ and we assume reducing conditions. 
    We show HEOM rates for varying truncation $T$ and one Matsubara mode $M=1$. For $T=1$ excellent agreement is achieved between HEOM rates and Redfield as expected, but converged HEOM rates at $T=4$ are in stark disagreement with Redfield.  
    For the converged results we see a reduced rate for the frequency domain that Redfield predicts a large rate, and in a) we observe a peak at half a frequency of the maximum indicating that two-phonon harmonic process are assisting population transfer.
    These results are not averaged over disorder.}
    \label{fig:rates_ubo_scan}
\end{figure*}

There are two observable indications of the redox mechanism in FMO that a theory should be able to reproduce. 
First and foremost is the small change in linear absorption spectra, which effectively tightly bounds the possible change in electronic parameters. 
The second is the relative change in the transfer pathways measured in the 2DES experiments. 
Figure \ref{Fig:linear_spectra}(a) shows experimental absorption spectra obtained at 77K, demonstrating that oxidation causes a red shift of the spectra while simultaneously maintaining its characteristic shape to a large extent. 
We simulated the absorption spectra generated by the proposed redox model with HEOM, where shifts to the energies of sites II, III, and IV are introduced, and the results are plotted in Fig.\ \ref{Fig:linear_spectra}(b). 
The oxidised model neither has an appropriate qualitative nor quantitative agreement with the measured absorption, particularly owing to its failure to recover its characteristic 825 nm peak. 
We further see that the proposed site shifts result in blue-shifted oxidation spectra, which is contrary to the experiment. 
An additional difference is the increased broadening of the linear spectra due to the presence of the UBO mode.
When considered with HEOM this mode acts to increase transfer rates between all excitonic states, not only those close to resonance as in the Redfield case as Table \ref{tab:main_rates_table} shows, which leads to an overall increase in the homogeneous broadening of the spectra.

Furthermore, the rates in Table \ref{tab:main_rates_table} show that while the model recovers the observed rates when simulated using Redfield theory, when using HEOM we fail to reproduce the result. 
The inclusion of the underdamped mode does not have the effect of precisely increasing a particular transition in the HEOM case, but rather acts broadly to increase rates for all transitions. 
This indicates that the vibrational influence approximated by the Redfield approach is inconsistent with HEOM, and that non-perturbative approaches are vital in accurately capturing such phenomena even phenomenologically. 
We analyse this in more detail in the next section where we see how various rates vary as a function of the UBO mode peak position.

\section{Study of vibrationally enhanced transfer rates}\label{sec:env_model}
We now analyse how generalised excitonic transfer rates derived within the HEOM formalism behave as a function of the underdamped vibrational mode peak position $\Omega$. 
To do so, we utilise the Drude-Lorentz form of the spectral density as our low-energy phonon background,
\begin{equation}\label{eq:drude_lorentz_spectral_density}
    J_{p}(\omega) = 2\lambda_{p} \gamma_p \frac{\omega}{\omega^2 + \gamma_p^2},
\end{equation}
with $\lambda_{p} = 35$ cm$^{-1}$ and $\gamma_p = 106$ cm$^{-1}$, which were chosen to coincide with the parameters that have been studied using HEOM before \cite{Hein2012Feb}. 
This is computationally more efficient to treat with HEOM, thus enabling a more thorough exploration of parameter regimes. 
In addition, we have a discrete vibrational mode as before (Eq.\ (\ref{eq:ubo_spectral_density})), with a smaller reorganisation energy $\lambda_u = 10$ cm$^{-1}$ and $\gamma_u = 8$ cm$^{-1}$. 
The full spectral density is now site-independent and of the form
\begin{equation}
    J(\omega) = J_{p}(\omega) + J_{u}(\omega).
\end{equation}

In Fig.\ \ref{fig:rates_ubo_scan}, we plot the rates $W_{4\to1}$, $W_{4\to2}$, $W_{2\to1}$ as a function of the peak position $\Omega$ for different levels of hierarchy truncation, which is contrasted with the corresponding Redfield result. 
Note that all HEOM results include one Matsubara mode.
For a truncation depth $T=1$ there is excellent quantitative agreement with Redfield, as expected \cite{Trushechkin2019Oct}, and both theories show a clear increase in transfer as a result of the resonant vibration. 
When the truncation is increased, such that converged results are obtained for $T=4$, this effect diminishes considerably. 
We were also able to capture additional structure in the non-perturbative results, in contrast to the single peak in Redfield. 
Most notable is a peak associated to a two-phonon harmonic process assisting transfer observed in the $W_{4\to1}$ results. 

\begin{figure}
    \centering
    \includegraphics[width=\linewidth]{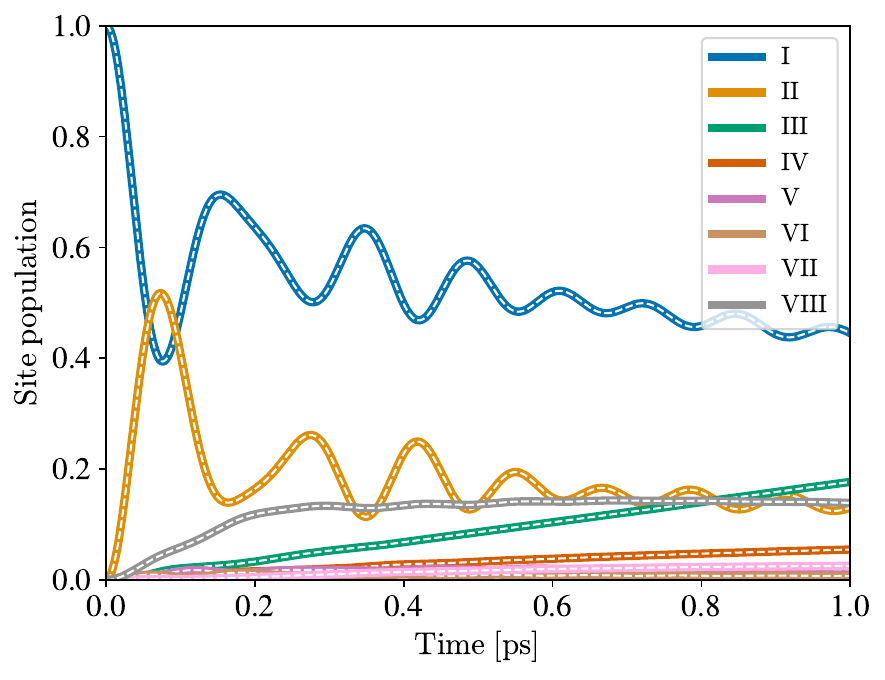}
    \caption{Comparison of site population dynamics computed using our HEOM implementation (solid lines) to QuTip \cite{Johansson2013Apr, Lambert2023Mar} (white dashed lines). 
    This set of dynamics corresponds to the model discussed in Section \ref{sec:env_model} for reduced FMO. 
    Calculations have been carried out with a truncation depth $T = 4$ and a single Matsubara mode $M = 1$, assuming a temperature of $77$K. }
    \label{fig:FMO_heom_qutip}
\end{figure}

As a result we chose the mode peak position for the reduced FMO model to be $\Omega = 260$ cm$^{-1}$ in order to assist the $4\to1$ pathway, whereas for the oxidised FMO model we have it at $\Omega = 150$ cm$^{-1}$ in order to assist the $4\to2$ transition. 
To further benchmark our HEOM implementation, we demonstrate in Fig.\ \ref{fig:FMO_heom_qutip} that it generates identical dynamics as the open-source Quantum Toolbox in Python (QuTip) implementation of HEOM \cite{Johansson2013Apr, Lambert2023Mar} for the reduced FMO model.
For oxidising conditions, we also introduce shifts $-10$ cm$^{-1}$, $-25$ cm$^{-1}$, $-25$ cm$^{-1}$ to sites II, III, and IV, respectively, in order to emulate the experimental absorption and 2DES results qualitatively. 
We refer to this combined negative site energy shift and vibrational mode shift as the altered environment (AE) model. 
Fig. \ref{Fig:linear_spectra}(c) shows the resulting absorption spectra averaged over disorder, which demonstrates a fidelity to the experimental spectra. 
This indicates that the site energy shifts compatible with reproducing the absorption spectra should be smaller in magnitude in comparison to what has been applied for Fig.\ \ref{Fig:linear_spectra}(b).
Excitonic transfer rates in Table \ref{tab:main_rates_table} corresponding to the AE model do not however capture the experimentally observed pathway tuning. 
This points to a potential incongruence between the experimental results and the theoretical model we are applying. 

\begin{figure*}
    \centering
    \includegraphics[width=\linewidth]{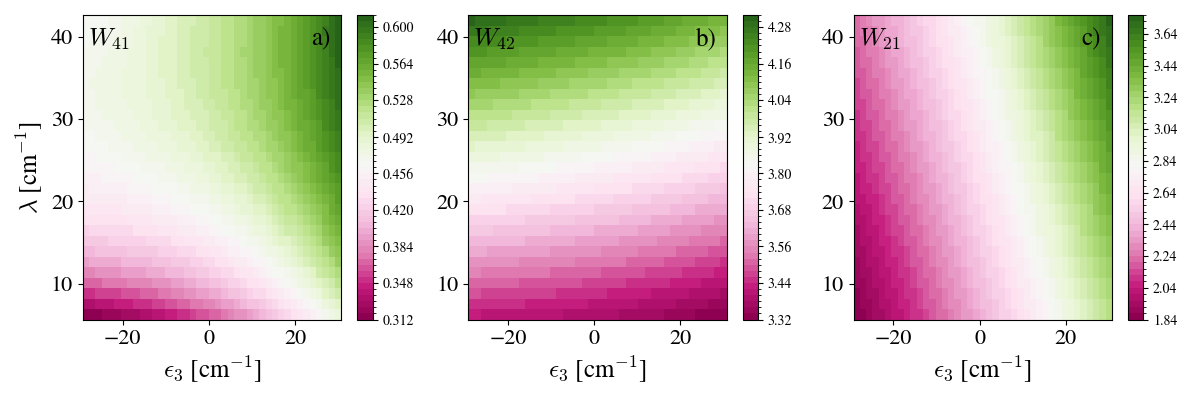}
    \caption{Change in HEOM exciton transfer rates a) $W_{4\to1}$, b) $W_{4\to2}$ and c) $W_{2\to1}$ with site energy $\epsilon_3$ and UBO reorganisation energy for each site. Environment describes a Drude-Lorentz background with $\lambda_{p} = 35 $ cm$^{-1}$ and $\gamma_{p} = 106$ cm$^{-1}$. Shown for $\Omega = 260$ cm$^{-1}$ and $\gamma_u = 8$ cm$^{-1}$. All other site energy shifts taken as zero. No disorder averaging is shown here.}
    \label{fig:ratesmap_energies}
\end{figure*}

\section{Redox mechanism modelled by site energy shifting}\label{sec:redox_mechanism}
In this section we continue to use the model introduced in Section \ref{sec:env_model} and explore how site shifts affect EET pathways.
In Fig.\ \ref{fig:ratesmap_energies} we show how the relevant rates are altered for varying site energies and vibrational reorganization energies $\lambda_u$. 
We see that over the range of site energies studied, which we choose to encompass the maximum range concomitant with a consistent absorption spectrum, we do not observe a significant enough change in exciton transfer rates to account for the redox mechanism except at large UBO reorganisation energies, which would in turn drastically broaden the absorption spectra.

Figure \ref{fig:ratesmap_energies} demonstrates an opposing tendency of $W_{4\to2}$ compared with $W_{4\to1}$ and $W_{2\to1}$ as $\epsilon_3$ and $\lambda_u$ are altered, which qualitatively matches the trend in the experimental data.
We thus observe that some selectivity of the transfer rates is observable from the site energy shifts. 
However, the magnitude of the rate change from a shift in the site energy alone is too small to account for experimental observations. 
In Appendix \ref{app:rate_tabels} we show similar plots for sites II and IV, again observing results that agree qualitatively with the direction of rate shifts seen in experiment\cite{Higgins2021Mar}, but with a reduced magnitude.
We observe that the combined small site energy shift and vibrational frequency shift are unable to reproduce the quantitative rate changes observed in experiment, yet capture the general trend that is seen in the redox mechanism. 
Note that our results do not rule out a site shift model for oxidation generally, but simply that the current Hamiltonian and environment do not reproduce the correct rates upon the introduction of sufficiently small shifts in the considered sites to reproduce the observed spectra. 
An altered Hamiltonian for the reduced form may reproduce the relevant oxidised rates upon site energy shifts alone. 
We speculate that a refined Hamiltonian based on fitting to non-perturbative methods may remedy this shortcoming; however, it is also possible that inter-site couplings may additionally change upon oxidation. 
Such a change in couplings may lead to differing excitonic delocalisation without a dramatic shift in the linear spectra \cite{Kulkarni2024Jul}, and thus may also contribute to the observed pathway modulation.

\section{Conclusions}\label{sec:conclusions}
In this work we theoretically investigated the redox-sensitive exciton transfer rates in the FMO complex using a reformalisation of HEOM into a memory kernel. 
The FMO complex has been experimentally observed to attenuate the direct exciton $4\to1$ pathway in comparison to the indirect exciton $4\to 2\to1$ pathway under oxidative conditions \cite{Higgins2021Mar}. 
This was proposed to be a result of the removal of a resonance between a vibrational energy scale and a relevant exciton energy gap due to site energy shifts upon oxidation.
We have shown that the originally proposed model is inconsistent with linear absorption spectra. 
By deriving generalised rates and spectra within the HEOM framework, we have also shown that site energy shifts alone do not yield both consistent linear spectra and transfer rates with experimental observation.
We then systematically investigated a model that includes both site energy shifts and change in the vibrational frequency of an environmental mode that assists exciton transfer. 
With this framework, we were able to observe changes in exciton transfer rates which qualitatively reproduced the experimental observations for reduced and oxidised FMO.
However, a quantitative agreement with the large changes in the rates observed in experiment was not recovered.

Furthermore, we observed that the differences between the effective exciton population transfer rates with Redfield and HEOM are not only quantitative but also qualitative.
Notably, we observe a much lower targeted increase in transition rates via a vibrational resonance condition with HEOM than with Redfield, and we observe significant structure in the the HEOM case that is absent in a Redfield approach. 
This structure, which manifests as rates that peak at different environmental frequencies, is due to multi-phonon processes contributing to population transfer which is captured by HEOM but not by Redfield. 
It is therefore clear that non-perturbative numerical techniques are critical in capturing the underlying mechanisms of exciton transfer in such systems due to the strong environmental coupling to a structured vibrational environment, that lead to non-Markovian behaviour. 

Further work is required to obtain a model that is sufficiently accurate to describe not only the red shift in oxidised FMO spectra, but also the quantitative exciton transfer rate changes in its reduced form. 
Reasons for the observed deviation may be that the employed Hamiltonian for reduced FMO, namely, model C by Kell \textit{et al.} \cite{Kell2016Aug}, was obtained by fitting a variety of experimental data according to second-order cumulant expansion and Redfield theory \cite{Adolphs2006Oct, Kell2016Aug}. 
Thus, it is not surprising that Redfield theory might capture the rates more closely for this model. 
A promising future line of research would be to obtain Hamiltonians from experimental data fitted using non-perturbative theories. 
These may better capture the role of the environment in determining the transfer rates in the system. 
For such a Hamiltonian, we note that oxidation could potentially be described by site shifts alone; although our analysis above suggests that in such a case, the system would require a highly sensitive dependence of the rates on shifts of the site energies, as there is only a small change in the observed linear spectra between these cases. 
These results lead us to suggest that a possible avenue to propose a more accurate model would be to revise the FMO Hamiltonian fitting done by Kell \textit{et al.} \cite{Kell2016Aug} but now within the HEOM or any other non-perturbative framework. 
Our work also motivates the use of first-principles approaches \cite{Kim2020Mar} to derive the electronic parameters for reduced and oxidised FMO.

\section*{Data Availability}
The Python code developed during this study and further data are available from the corresponding author upon reasonable request.

\section*{Acknowledgements}
H.Ó.G., C.N., and A.O-C thank the Gordon and Betty Moore Foundation for financial support (Grant GBMF8820). A.O-C and C.N.\ also acknowledge funding from the Engineering and Physical Sciences Research Council (EPSRC UK), Grant EP/V049011/1. J.S.H.\ received support from a National Research Council Postdoctoral Fellowship and G.S.E.\ acknowledges financial support from the U.S. Department of Energy, Basic Energy Science, through Award No. DE-SC0020131 and the National Science Foundation through QuBBE QLCI (NSF OMA-2121044). The authors acknowledge the use of the UCL Myriad High Performance Computing Facility.

\appendix

\section{Model C}\label{app:FMO}
The electronic Hamiltonian generated by Kell \textit{et al.} \cite{Kell2016Aug} is
\begin{widetext}
    \begin{equation}
        \hat{H}_S = 
        \begin{bmatrix}
            12405& -87 & 4.2 & -5.2 & 5.5 & -14 & -6.1 & 21 \\
            -87 & 12505 & 28 & 6.9 & 1.5 & 8.7 & 4.5 & 4.2 \\
            4.2 & 28 & 12150 & -54 & -0.2 & -7.6 & 1.2 & 0.6 \\
            -5.2 & 6.9 & -54 & 12300 & -62 & -16 & -51 & -1.3 \\
            5.5 & 1.5 & -0.2 & -62 & 12470 & 60 & 1.7 & 3.3 \\
            -14 & 8.7 & -7.6 & -16 & 60 & 12575 & 29 & -7.9 \\
            -6.1 & 4.5 & 1.2 & -51 & 1.7 & 29 & 12375 & -9.3 \\
            21 & 4.2 & 0.6 & -1.3 & 3.3 & -7.9 & -9.3 & 12430
        \end{bmatrix},
    \end{equation}
\end{widetext}
with elements in units of cm$^{-1}$. This model applies a log-normal distribution to its spectral density, which is of the form
\begin{equation}
    J(\omega) = \frac{S}{\sqrt{2\pi}\sigma}\omega e^{-\ln^2(\omega / \omega_c) / 2\sigma^2},
\end{equation}
where $S$ is the Huang-Rhys factor, $\sigma$ is the standard deviation, and $\omega_c$ is the frequency cut-off. 
For sites 1,2,4,5,6,7,8 we have $\omega_c = 45$ cm$^{-1}$, $\sigma = 0.85$, $S = 0.4\pi$, and for site 3 we have $\omega_c = 38$ cm$^{-1}$, $\sigma = 0.7$, $S = 0.4\pi$. 
Note that to convert from the convention we use, to the one used by Kell \textit{et al.} \cite{Kell2016Aug}, one must send $J(\omega)\rightarrow J(\omega) / (\pi\omega^2)$. 
Finally, for static disorder we have for sites 1,2,4,5,6,7,8 a full width at half maximum of 125 cm$^{-1}$, and for site 3 it is 75 cm$^{-1}$. 
The dipoles were obtained from the Protein Data Bank entry 3ENI \cite{Tronrud2009May}. 

\subsection{Reduced and oxidised models}
Higgins \textit{et al.} \cite{Higgins2021Mar} posit that an underdamped mode should be added to the spectral density of model C to match the experimental transfer rate data. It has the form
\begin{equation}
    J_g(\omega) = \frac{S_g}{\sqrt{2\pi}\sigma_g}\omega^2 e^{-\frac{(\omega - \Omega_g)^2}{2\sigma_g^2}},
\end{equation}
where $S_g = 0.375$, $\sigma_g = 4.25$ cm$^{-1}$, and $\Omega_g = 260$ cm$^{-1}$. 
This modification of the spectral density was applied to both the reduced and oxidised model.

They furthermore introduce shifts to the site-energies $\Delta\epsilon_2 = $ 40 cm$^{-1}$, $\Delta\epsilon_3 = $ 120 cm$^{-1}$, $\Delta\epsilon_4 = $ 70 cm$^{-1}$, such that $\braket{n\vert \hat{H}_{\text{Oxi}}\vert n} = \braket{n\vert \hat{H}\vert n} + \Delta\epsilon_n$ for $n = 2, 3, 4$, which then serves as the oxidised model of FMO.

\section{Absorption spectra}\label{app:linear_absorption}
We use the standard Wiener-Khinchin theorem to simulate the absorption spectrum of FMO. The absorption spectrum is given as the Fourier transform of the total transition dipole auto-correlation function
\begin{equation}\label{eq:absorption_spectra_formal}
    A(\omega) = \sum_{p = x,y,z} \int_{-\infty}^\infty\text{d}t \braket{\hat{\mu}_p(t)\hat{\mu}_p(0)} e^{i\omega t},
\end{equation}
where the average is taken with respect to the zero excitation state of the FMO, denoted as $\rho_0 = \ket{0}\bra{0}$ (a stationary state), and we have $\hat{\mu}_p = \sum_i \mu_{p,i}\ket{i}\bra{0} + \text{h.c.}$ 
The HEOM yields a non-unitary expansion of the Feynman-Vernon influence functional such that the negative time component of Eq.\ \eqref{eq:absorption_spectra_formal} is problematic, because the backwards-in-time propagation of an irreversible process is non-physical. 
Eq.\ \eqref{eq:absorption_spectra_formal} is therefore rewritten formally as
\begin{equation}\label{eq:absorption_qrt_ready}
    A(\omega) = 2\sum_{p=x,y,z}\text{Re}\left[\int_0^\infty\text{d}t \braket{\hat{\mu}_p(t)\hat{\mu}_p(0)} e^{i\omega t}\right],
\end{equation}
where we have made use of the fact that $\braket{\hat{\mu}_p(0)\hat{\mu}_p(t)} = \braket{\hat{\mu}_p(t)\hat{\mu}_p(0)}^*$. In this form we now apply the quantum regression theorem, wherein the generator of the reduced system density matrix dynamics is assumed to be the generator of dynamics for the operator $\hat{\mu}_p\rho_0$ \cite{Carmichael1993}. Note that the quantum regression theorem calls for system-environment correlations to be discarded, a caveat that does not have any implications here, as $\hat{\mu}_p\rho_0$ will be uncorrelated to the environment. 
We then have
\begin{equation}
    A(\omega) = 2\sum_{p=x,y,z}\text{Re}\left[\int_0^\infty\text{d}t \text{Tr}\left(\hat{\mu}_p e^{\mathcal{L}t}\left[\hat{\mu}_p\rho_0\right]\right)e^{i\omega t}\right],
\end{equation}
where $\mathcal{L}$ is the generator of the reduced system density matrix dynamics, $\partial_t\rho = \mathcal{L}\rho$. 
We may formally integrate this expression to obtain
\begin{equation}
    A(\omega) = -2\sum_{p=x,y,z}\text{Re}\left[\text{Tr}\left(\hat{\mu}_p \frac{1}{\mathcal{L} + i\omega}\left[\hat{\mu}_p\rho_0\right]\right)\right],
\end{equation}
which is more convenient for our purposes because the problem of solving the equations of motion is then replaced by the problem of solving the operator equation
\begin{equation}
    (\mathcal{L} + i\omega)\hat{x}_{p,\omega} = \hat{\mu}_p\rho_0,
\end{equation}
such that the absorption spectrum may then finally be computed numerically as
\begin{equation}
    A(\omega) = -2\sum_{p=x,y,z}\text{Re}\left[\text{Tr}\left(\hat{\mu}_p \hat{x}_{p,\omega}\right)\right].
\end{equation}

\section{Redfield exciton rates}\label{app:redfield}
Redfield theory relies on the assumption that the coupling between system and environment is small.
The Redfield quantum master equation is
\begin{align}
    \partial_t \rho_{\alpha\beta} = -i \omega_{\alpha\beta} \rho_{\alpha\beta} + \sum_{\gamma,\delta}R_{\alpha\beta\gamma\delta} \rho_{\gamma\delta},
\end{align}
with $\rho_{\alpha\beta} = \langle \alpha |\rho |\beta\rangle$ where $\ket{\alpha}$,$\ket{\beta}$ are eigenstates of the Hamiltonian with energies $E_\alpha$, $E_\beta$ and $\omega_{\alpha\beta} = E_\alpha - E_\beta$. The rate of population transfer from $\rho_{\alpha\alpha}$ to $\rho_{\beta\beta}$ is given by the Redfield tensor element $R_{\beta\beta\alpha\alpha}$, where
\begin{align}
    R_{\alpha\beta\gamma\delta} = \Gamma_{\delta\beta\alpha\gamma} + \Gamma_{\gamma\alpha\beta\delta}^* - \delta_{\beta\delta} \sum_{\xi} \Gamma_{\alpha \xi\xi \gamma}  - \delta_{\alpha\gamma} \sum_{\xi} \Gamma_{\beta \xi\xi \delta}^*,
\end{align}
with $\Gamma_{\alpha\beta\gamma\delta} = \sum_{n} \langle \alpha|n\rangle \langle n|\beta\rangle \langle \gamma|n\rangle \langle n|\delta\rangle \tilde{C}_{n}(\omega_{\delta\gamma})$.
Here $\tilde{C}_{n}(\omega)$ is the half-sided Fourier transform of the correlation function
\begin{equation}
    \tilde{C}_{n}(\omega) = \int_0^\infty \text{d}t C_{n}(t)e^{i\omega t},
\end{equation}
which in turn is related to the spectral density $J_n(\omega)$ via
\begin{equation}
    C_{n}(t) = \frac{1}{\pi} \int_{-\infty}^{\infty} \text{d}\omega J_{n}(\omega) (1 + n(\omega))e^{-i\omega t},
\end{equation}
where we have extended the domain of the spectral density to negative frequencies via the prescription $J(\omega) = -J(-\omega)$ for $\omega < 0$, and introduced the Bose-Einstein distribution $n(\omega) = (e^{\beta\omega} - 1)^{-1}$. 
The exciton population transfer rate is then determined as
\begin{align}
    R_{\beta\beta\alpha\alpha} &= 2\text{Re}(\Gamma_{\alpha\beta\beta\alpha}) \\
    &= 2\sum_n \braket{\alpha\vert n}^2 \braket{\beta\vert n}^2 \text{Re}(\tilde{C}_{n}(\omega_{\alpha\beta})) \\
    &= 2(1 + n(\omega_{\alpha\beta}))\sum_n \braket{\alpha\vert n}^2 \braket{\beta\vert n}^2 J_n(\omega_{\alpha\beta})
\end{align}
where it is assumed that $\alpha\neq\beta$.

\section{HEOM expansion coefficients}\label{app:heom}
The HEOM approach to open quantum system dynamics is based on the assumption that the environmental correlation functions can be expressed in terms of the sum of exponentials
\begin{align}\label{eq:exp_corr_func}
    C_n(t) = \sum_k c_{n,k} e^{-\nu_{n,k} t},
\end{align}
where, for a Drude-Lorentz spectral density of the form given in Eq.\ \eqref{eq:drude_lorentz_spectral_density} with reorganisation energy $\lambda_n$ and damping $\gamma_n$ we have the coefficients
\begin{align}
    \nu_{n,0} & = \gamma_n, \\
    \nu_{n,k} & = 2\pi k / \beta, \\
    c_{n,0} & = \lambda_n\gamma_n\left(\cot\left(\frac{\beta\gamma_n}{2}\right) - i\right), \\
    c_{n,k} & = \frac{4\lambda_n\gamma_n}{\beta}\frac{\nu_{n,k}}{\nu_{n,k}^2 - \gamma_n^2},
\end{align}
where $k = 1, 2, 3\ldots$\, For an underdamped Brownian oscillator, the spectral density of the form given in Eq.\ \eqref{eq:ubo_spectral_density} with reorganisation energy $\lambda_n$, damping $\gamma_n$, and energy shift $\Omega_n$ we have the coefficients
\begin{align}
    \nu_{n,\pm} &= \frac{\gamma_n}{2} \pm i\sqrt{\Omega_n^2 - \frac{\gamma_n^2}{4}}, \\
    \nu_{n,k} &= 2\pi k / \beta, \\
    c_{n,\pm} &= \pm\frac{\lambda_n\Omega_n^2}{2\omega_n^\prime}\left(1 + i\cot\left(\frac{\beta\gamma_{n,\pm}}{2}\right)\right), \\
    c_{n,k} &= -\frac{4\lambda_n\gamma_n}{\beta}\frac{\Omega_n^2\nu_{n,k}}{(\gamma^2_{n,+}-\nu_{n,k}^2)(\gamma^2_{n,-}-\nu_{n,k}^2)}, 
\end{align}
with $k = 1, 2, 3\ldots$ and $\omega_n^\prime = \sqrt{\Omega_n^2 - \frac{\gamma_n^2}{4}}$.

\section{Calculated rates for each model with Redfield and HEOM formalism}\label{app:rate_tabels}
Here, for completion, we report all calculated exciton transfer rates for each model with both Redfield and HEOM where relevant. We show that, comparing Table \ref{table:redfield_log_normal} and Table \ref{table:redfield_ubo_fit}, the fit of the UBO environment within the Redfield approximation does not significantly alter any rates. This indicates that whilst the log-normal form may have preferable low energy behaviour, the UBO is a good model that enables numerically exact calculations. We may then compare directly the Redfield and HEOM approaches via Tables \ref{table:redfield_ubo_fit} and \ref{table:heom_ubo_fit}, respectively. In this case we see a significant change in the effect of the second UBO mode, which in the HEOM calculation does not act to significantly alter the 4-1 rate as it does in the Redfield calculation. Finally, in Table \ref{table:heom_alternate_model_v2}, we show the approximately three-fold decrease in the transition rate of the 4-1 transition, which is captured by a shift in the second UBO mode frequency.

We show additional HEOM rates colour plots in Fig.\ \ref{fig:ratesmap_energies2} below. We observe very little change in rates with site energy $\epsilon_2$, however a small increase in $\epsilon_4$ is observed to decrease $W_{4\to1}$, except for at very high UBO reorganisation energies where increasing $\epsilon_4$ slightly decreases $W_{4\to1}$. Increasing $\epsilon_4$ is also seen to increase $W_{4\to2}$, and decrease $W_{2\to1}$, for all values of $\lambda$. We note that in each case, however, the change in rates are small for a change in site energy alone within the limits required to reproduce the observed linear spectra.

\begin{table*}[]
    \caption{Tables of mean Redfield rates, in units of ps$^{-1}$, for a) the log-normal density or Ref. \cite{Kell2016Aug} and b) the inclusion of a Gaussian peak as in Ref. \cite{Higgins2021Mar}. Cells highlighted in yellow refer to the rates of interest: $W_{4\to1}$, $W_{4\to2}$, and $W_{2\to1}$.}
    \begin{minipage}{.5\linewidth}
        \begin{tabular}{|c|c|c|c|c|c|c|c|c|}\hline
            \multicolumn{9}{|c|}{Without Gaussian mode}\\\hline
            \hline
             & 1 & 2 & 3 & 4 & 5 & 6 & 7 & 8 \\
            \hline
            1 &  & \cellcolor{yellow!55}1.96 & 0.20 & \cellcolor{yellow!55}0.14 & 0.07 & 0.05 & 0.02 & 0.01 \\
            \hline
            2 & 0.22 &  & 0.94 & \cellcolor{yellow!55}2.08 & 0.98 & 0.58 & 0.21 & 0.11 \\
            \hline
            3 & 0.01 & 0.22 &  & 0.99 & 1.14 & 0.63 & 1.62 & 0.26 \\
            \hline
            4 & 0.00 & 0.32 & 0.39 &  & 1.33 & 1.00 & 0.79 & 0.43 \\
            \hline
            5 & 0.00 & 0.08 & 0.32 & 0.67 &  & 0.98 & 0.85 & 1.03 \\
            \hline
            6 & 0.00 & 0.02 & 0.07 & 0.29 & 0.53 &  & 1.14 & 1.95 \\
            \hline
            7 & 0.00 & 0.00 & 0.04 & 0.04 & 0.10 & 0.34 &  & 1.25 \\
            \hline
            8 & 0.00 & 0.00 & 0.00 & 0.01 & 0.05 & 0.14 & 0.55 &  \\
            \hline
        \end{tabular}
    \end{minipage}%
    \begin{minipage}{.5\linewidth}
        \begin{tabular}{|c|c|c|c|c|c|c|c|c|}\hline
            \multicolumn{9}{|c|}{With Gaussian mode}\\\hline
            \hline
             & 1 & 2 & 3 & 4 & 5 & 6 & 7 & 8 \\
            \hline
            1 &  & \cellcolor{yellow!55}2.00 & 0.79 & \cellcolor{yellow!55}2.09 & 1.19 & 0.32 & 0.02 & 0.01 \\
            \hline
            2 & 0.22 &  & 0.96 & \cellcolor{yellow!55}2.08 & 1.10 & 1.29 & 2.64 & 0.56 \\
            \hline
            3 & 0.01 & 0.23 &  & 0.98 & 1.15 & 0.67 & 5.69 & 3.56 \\
            \hline
            4 & 0.02 & 0.31 & 0.39 &  & 1.32 & 1.00 & 1.12 & 2.84 \\
            \hline
            5 & 0.01 & 0.08 & 0.32 & 0.66 &  & 0.97 & 0.87 & 1.79 \\
            \hline
            6 & 0.00 & 0.03 & 0.07 & 0.30 & 0.53 &  & 1.14 & 2.14 \\
            \hline
            7 & 0.00 & 0.02 & 0.07 & 0.04 & 0.10 & 0.34 &  & 1.25 \\
            \hline
            8 & 0.00 & 0.00 & 0.03 & 0.03 & 0.06 & 0.14 & 0.55 &  \\
            \hline
        \end{tabular}
    \end{minipage} 
    \label{table:redfield_log_normal}
\end{table*}

\begin{table*}[]
    \caption{Tables of mean Redfield rates, in units of ps$^{-1}$, for the UBO best-fit as shown in Figure \ref{fig:ln_vs_fit} for a) the log-normal density or Ref. \cite{Kell2016Aug} and b) the inclusion of a Gaussian peak as in Ref. \cite{Higgins2021Mar}. Cells highlighted in yellow refer to the rates of interest: $W_{4\to1}$, $W_{4\to2}$, and $W_{2\to1}$.}
    \begin{minipage}{.5\linewidth}
        \begin{tabular}{|c|c|c|c|c|c|c|c|c|}\hline
            \multicolumn{9}{|c|}{Without second UBO.}\\\hline
            \hline
             & 1 & 2 & 3 & 4 & 5 & 6 & 7 & 8 \\
            \hline
            1 &  & \cellcolor{yellow!55}1.95 & 0.21 & \cellcolor{yellow!55}0.14 & 0.07 & 0.05 & 0.02 & 0.01 \\
            \hline
            2 & 0.22 &  & 1.11 & \cellcolor{yellow!55}2.07 & 0.99 & 0.59 & 0.22 & 0.11 \\
            \hline
            3 & 0.01 & 0.36 &  & 1.21 & 1.16 & 0.64 & 1.65 & 0.27 \\
            \hline
            4 & 0.00 & 0.31 & 0.58 &  & 1.81 & 1.00 & 0.81 & 0.44 \\
            \hline
            5 & 0.00 & 0.08 & 0.33 & 1.02 &  & 1.51 & 0.87 & 1.05 \\
            \hline
            6 & 0.00 & 0.02 & 0.07 & 0.30 & 0.92 &  & 1.30 & 1.98 \\
            \hline
            7 & 0.00 & 0.00 & 0.04 & 0.04 & 0.10 & 0.45 &  & 1.49 \\
            \hline
            8 & 0.00 & 0.00 & 0.00 & 0.01 & 0.05 & 0.14 & 0.70 &  \\
            \hline
        \end{tabular}
    \end{minipage}%
    \begin{minipage}{.5\linewidth}
        \begin{tabular}{|c|c|c|c|c|c|c|c|c|}\hline
            \multicolumn{9}{|c|}{With second UBO.}\\\hline
            \hline
             & 1 & 2 & 3 & 4 & 5 & 6 & 7 & 8 \\
            \hline
            1 &  & \cellcolor{yellow!55}2.22 & 0.95 & \cellcolor{yellow!55}2.54 & 1.43 & 0.44 & 0.03 & 0.01 \\
            \hline
            2 & 0.24 &  & 1.11 & \cellcolor{yellow!55}2.13 & 1.26 & 1.63 & 3.17 & 0.68 \\
            \hline
            3 & 0.02 & 0.35 &  & 1.22 & 1.17 & 0.74 & 7.37 & 3.98 \\
            \hline
            4 & 0.02 & 0.32 & 0.58 &  & 1.80 & 1.02 & 1.44 & 3.35 \\
            \hline
            5 & 0.01 & 0.09 & 0.33 & 1.01 &  & 1.48 & 0.95 & 2.07 \\
            \hline
            6 & 0.00 & 0.04 & 0.07 & 0.30 & 0.91 &  & 1.31 & 2.50 \\
            \hline
            7 & 0.00 & 0.03 & 0.10 & 0.05 & 0.11 & 0.44 &  & 1.48 \\
            \hline
            8 & 0.00 & 0.00 & 0.03 & 0.04 & 0.07 & 0.16 & 0.69 &  \\
            \hline
        \end{tabular}
    \end{minipage} 
    \label{table:redfield_ubo_fit}
\end{table*}

\begin{table*}[]
    \caption{Tables of mean HEOM rates, in units of ps$^{-1}$, for the UBO best-fit shown in Figure \ref{fig:ln_vs_fit} to the spectral density of Refs. \cite{Kell2016Aug, Higgins2021Mar}. Cells highlighted in yellow refer to the rates of interest: $W_{4\to1}$, $W_{4\to2}$, and $W_{2\to1}$.}
    \begin{minipage}{.5\linewidth}
        \begin{tabular}{|c|c|c|c|c|c|c|c|c|}\hline
            \multicolumn{9}{|c|}{Reduced (PNAS)}\\\hline
            \hline
             & 1 & 2 & 3 & 4 & 5 & 6 & 7 & 8 \\
            \hline
            1 &  & \cellcolor{yellow!55}4.41 & 0.79 & \cellcolor{yellow!55}0.67 & 0.41 & 0.32 & 0.22 & 0.06 \\
            \hline
            2 & 2.18 &  & 2.39 & \cellcolor{yellow!55}4.86 & 2.97 & 2.23 & 0.99 & 0.80 \\
            \hline
            3 & 0.29 & 1.76 &  & 2.53 & 2.46 & 1.69 & 5.43 & 1.16 \\
            \hline
            4 & 0.24 & 3.10 & 2.08 &  & 3.61 & 2.11 & 2.41 & 1.74 \\
            \hline
            5 & 0.15 & 1.61 & 1.77 & 3.14 &  & 3.03 & 2.21 & 3.40 \\
            \hline
            6 & 0.10 & 1.05 & 0.99 & 1.56 & 2.68 &  & 2.75 & 5.52 \\
            \hline
            7 & 0.06 & 0.40 & 2.79 & 1.28 & 1.32 & 2.05 &  & 3.11 \\
            \hline
            8 & 0.03 & 0.23 & 0.50 & 0.77 & 1.72 & 3.01 & 2.57 &  \\
            \hline
        \end{tabular}
    \end{minipage}%
    \begin{minipage}{.5\linewidth}
        \begin{tabular}{|c|c|c|c|c|c|c|c|c|}\hline
            \multicolumn{9}{|c|}{Oxidised (PNAS)}\\\hline
            \hline
             & 1 & 2 & 3 & 4 & 5 & 6 & 7 & 8 \\
            \hline
            1 &  & \cellcolor{yellow!55}4.37 & 2.04 & \cellcolor{yellow!55}2.01 & 1.45 & 1.46 & 0.60 & 0.22 \\
            \hline
            2 & 2.73 &  & 2.76 & \cellcolor{yellow!55}4.09 & 2.32 & 1.94 & 1.96 & 0.86 \\
            \hline
            3 & 1.07 & 2.29 &  & 3.32 & 3.42 & 1.91 & 3.50 & 1.58 \\
            \hline
            4 & 0.98 & 2.86 & 2.72 &  & 3.69 & 2.94 & 2.21 & 1.89 \\
            \hline
            5 & 0.66 & 1.40 & 2.47 & 3.17 &  & 3.24 & 2.35 & 2.63 \\
            \hline
            6 & 0.60 & 0.99 & 1.13 & 2.16 & 2.80 &  & 3.07 & 5.25 \\
            \hline
            7 & 0.21 & 0.87 & 1.74 & 1.15 & 1.37 & 2.06 &  & 3.50 \\
            \hline
            8 & 0.09 & 0.29 & 0.68 & 0.86 & 1.33 & 2.90 & 2.99 &  \\
            \hline
        \end{tabular}
    \end{minipage} 
 \label{table:heom_ubo_fit}
\end{table*}

\begin{table*}[]
    \caption{FMO Oxidation Model 2. Reduced model has second UBO at 260 cm$^{-1}$, oxidised has second UBO at 150 cm$^{-1}$. Rates are given in units of ps$^{-1}$. Cells highlighted in yellow refer to the rates of interest: $W_{4\to1}$, $W_{4\to2}$, and $W_{2\to1}$.}
    \begin{minipage}{.5\linewidth}
        \begin{tabular}{|c|c|c|c|c|c|c|c|c|}\hline
            \multicolumn{9}{|c|}{Reduced FMO Model 2}\\\hline
            \hline
             & 1 & 2 & 3 & 4 & 5 & 6 & 7 & 8 \\
            \hline
            1 &  & \cellcolor{yellow!55}2.60 & 0.48 & 0.\cellcolor{yellow!55}45 & 0.28 & 0.21 & 0.10 & 0.04 \\
            \hline
            2 & 0.48 &  & 0.98 & \cellcolor{yellow!55}2.03 & 1.25 & 0.99 & 0.55 & 0.37 \\
            \hline
            3 & 0.06 & 0.38 &  & 1.01 & 1.04 & 0.72 & 2.62 & 0.63 \\
            \hline
            4 & 0.03 & 0.47 & 0.54 &  & 1.39 & 0.87 & 1.10 & 0.88 \\
            \hline
            5 & 0.02 & 0.19 & 0.36 & 0.89 &  & 1.16 & 0.92 & 1.51 \\
            \hline
            6 & 0.01 & 0.11 & 0.13 & 0.33 & 0.79 &  & 1.1 & 2.34 \\
            \hline
            7 & 0.01 & 0.03 & 0.22 & 0.12 & 0.18 & 0.47 &  & 1.21 \\
            \hline
            8 & 0.00 & 0.02 & 0.04 & 0.07 & 0.19 & 0.37 & 0.72 &  \\
            \hline
        \end{tabular}
    \end{minipage}%
    \begin{minipage}{.5\linewidth}
        \begin{tabular}{|c|c|c|c|c|c|c|c|c|}\hline
            \multicolumn{9}{|c|}{Oxidised FMO Model 2}\\\hline
            \hline
             & 1 & 2 & 3 & 4 & 5 & 6 & 7 & 8 \\
            \hline
            1 &  & \cellcolor{yellow!55}1.98 & 0.39 & \cellcolor{yellow!55}0.30 & 0.17 & 0.14 & 0.08 & 0.03 \\
            \hline
            2 & 0.30 &  & 1.01 & \cellcolor{yellow!55}2.73 & 1.68 & 1.25 & 0.54 & 0.34 \\
            \hline
            3 & 0.04 & 0.40 &  & 1.17 & 1.28 & 0.94 & 2.43 & 0.48 \\
            \hline
            4 & 0.02 & 0.64 & 0.58 &  & 1.50 & 1.06 & 1.03 & 0.78 \\
            \hline
            5 & 0.01 & 0.26 & 0.41 & 0.94 &  & 1.21 & 1.00 & 1.67 \\
            \hline
            6 & 0.01 & 0.14 & 0.15 & 0.39 & 0.81 &  & 1.28 & 2.89 \\
            \hline
            7 & 0.00 & 0.03 & 0.20 & 0.13 & 0.21 & 0.56 &  & 1.18 \\
            \hline
            8 & 0.00 & 0.02 & 0.03 & 0.07 & 0.23 & 0.45 & 0.61 &  \\
            \hline
        \end{tabular}
    \end{minipage} 
    \label{table:heom_alternate_model_v2}
\end{table*}

\begin{figure*}
    \centering
    \includegraphics[width=0.99\linewidth]{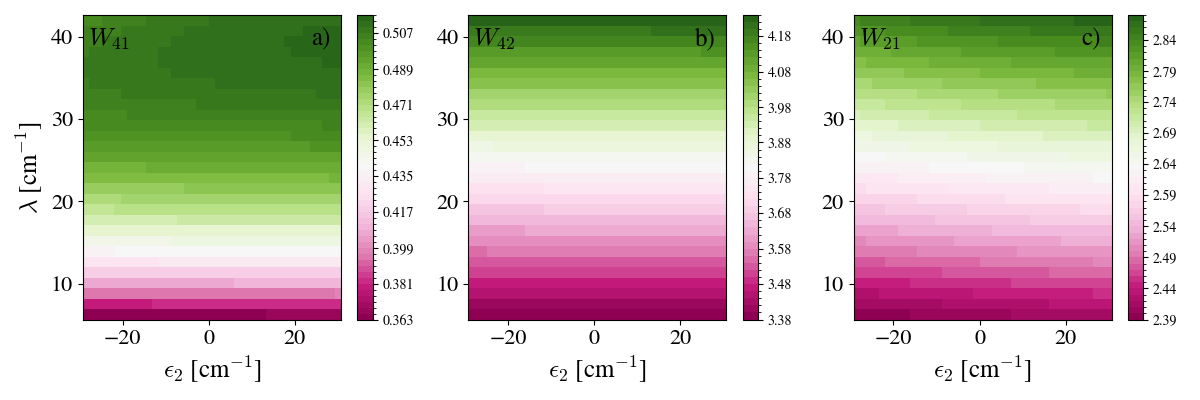}\\
        \includegraphics[width=0.99\linewidth]{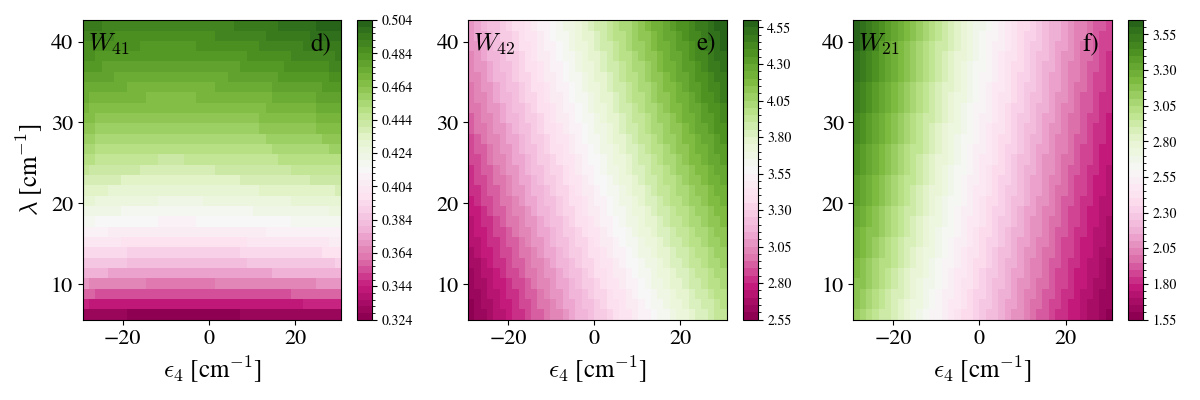}

    \caption{Change in HEOM exciton transfer rates (in units of ps$^{-1}$) with site energy a)-c)$\epsilon_2$ and d)-f) $\epsilon_4$ and UBO reorganisation energy. Environment describes a DL background with $\lambda_{DL} = 35 $ cm$^{-1}$ and $\Omega_{DL} = 106.1$ cm$^{-1}$. Shown for $\Omega = 315$ cm$^{-1}$, $\gamma = 7.9$ cm$^{-1}$. All other site energy shifts taken as zero.}
    \label{fig:ratesmap_energies2}
\end{figure*}

\bibliography{fmo_paper.bib}

\end{document}